\documentclass{webofc}
\usepackage[varg]{txfonts}   % Web of Conferences font
\woctitle{Fifteenth International Symposium on Capture Gamma-Ray
Spectroscopy and Related Topics (CGS15)}
\begin{document}
\title{Isospin transfer modes in exotic nuclei}

\author{Elena Litvinova\inst{1,2}\fnsep\thanks{\email{elena.litvinova@wmich.edu}}
}

\institute{Department of Physics, Western Michigan University,
Kalamazoo, MI 49008-5252, USA \and
           National Superconducting
Cyclotron Laboratory, Michigan State University, East Lansing, MI
48824-1321, USA
%
%\and Physics Department, Faculty of Science, University of Zagreb,
%Croatia
%                \and
%                Physik-Department der Technischen Universit\"at
%M\"unchen, D-85748 Garching, Germany
          }
\abstract{%
This work presents an approach to nuclear spin-isospin response,
which is capable of describing the overall strength distribution up
to high excitation energies, including the fine structure of the
low-lying strength, and resolving the long-standing quenching
problem. The model is a covariant realization of the nuclear field
theory and based on the self-consistent extensions of the covariant
energy density functional (CEDF) theory.
%The effective one-boson
%exchange interaction spans effective mesons and emerging collective
%modes. While heavy mesons are treated as classical fields, the
%low-lying collective phonons are included non-perturbatively in the
%time-blocking approximation. Thus, the covariant spin-isospin
%response theory is advanced to the inclusion of temporal and spatial
%non-localities. The approach based on a few parameters of the CEDF
%provides a high-quality description of nuclear excitation spectra in
%both neutral and charge-exchange channels.
Results of the recent
calculations for spin-isospin response of ordinary and exotic
medium-mass nuclei are presented and discussed. }
\maketitle
\section{Introduction}
\label{intro} The spin-isospin response associated with finite spin
and isospin transfer is one of the most important properties of
nuclei. This type of response provides information about a variety
of weak interaction processes, such as ordinary and double
beta-decay, electron capture, neutrino capture and scattering on
nuclei and in stars. Lately, the models, which are commonly used to
describe nuclear spin-isospin response, namely Quasiparticle Random
Phase Approximation and Shell Model, have advanced considerably.
However, a self-consistent approach, which can simultaneously
reproduce data on the overall strength distribution up to high
excitation energy, quenching, and on the fine structure of the
low-lying strength, is still a challenge.

This paper presents the spin-isospin version of the relativistic
time blocking approximation (RTBA) as a model, which can pretend to
account for all these effects. The spin-isospin RTBA (or
proton-neutron RTBA, pn-RTBA) has been developed and applied for the
first time to the spin-dipole resonance (SDR) in Ref. \cite{MLVR.12}
and, recently, to the Gamow-Teller resonance (GTR) in Ref.
\cite{LBFMZ.14}. Here, this approach is discussed in light of its
ability to describe unstable nuclear systems and to resolve the
long-standing problem of quenching of the GTR.

\section{Nuclear spin-isospin response: damping effects}
\label{sec-1}
%For bibliography use \cite{RefJ}
%\subsection{Subsection title}
%\label{sec-2}
%Don't forget to give each section, subsection, subsubsection, and
%paragraph a unique label (see Sect.~\ref{sec-1}).

The RTBA for the spin-isospin channel is based on the extended CEDF
theory \cite{LR.06,L.12} and constructed similarly to the R(Q)TBA
for the neutral channel developed earlier \cite{LRT.07,LRT.08}, its
non-relativistic version \cite{LT.07,TSG.07} and applications
\cite{E.10,E.12,M.12}. The only limitation so far is that the
spin-isospin RTBA does not include pairing correlations, but such a
generalization is straightforward and well underway. The
calculations presented below are performed in the following three
steps: (i) a relativistic mean field (RMF) solution is obtained by
minimization of the covariant density functional with NL3
parametrization \cite{NL3}, (ii) phonon spectrum and coupling
vertices for the phonons with $J^{\pi} = 2^+, 3^-, 4^+, 5^-, 6^+$
are obtained by the self-consistent relativistic RPA (RRPA)
\cite{RMGVWC.01} and (iii) the Bethe-Salpeter equation is solved for
the proton-neutron response function:
\begin{equation}
R(\omega) = {\tilde R}^{0}(\omega) + {\tilde R}^{0}(\omega)
W(\omega)R(\omega).
\end{equation}
Here ${\tilde R}^{0}(\omega)$ is the propagator of two uncorrelated
proton and neutron quasiparticles in the static mean field and the
integral part contains the in-medium nucleon-nucleon interaction
$W(\omega)$. The active channels of the two-body interaction
$W(\omega)$ allowing for spin-flip and isospin-flip include the
following static terms and the term depending on the frequency
$\omega$:
\begin{equation}
W(\omega) = V_{\rho} + V_{\pi} + V_{\delta\pi} + \Phi(\omega) -
\Phi(0). \label{inter}
\end{equation}
$V_{\rho}$ and $V_{\pi}$ are the finite-range $\rho$-meson and the
$\pi$-meson exchange interactions, respectively. They are derived
from the covariant energy density functional and read
\cite{PNVR.04}:
\begin{equation}
V_{\rho}(1,2) =
g_{\rho}^2{\vec\tau}_1{\vec\tau}_2(\beta\gamma^{\mu})_1(\beta\gamma_{\mu})_2
D_{\rho}({\bf r}_1,{\bf r}_2), \ \ \ \ \ \ \ \   V_{\pi}(1,2) = -
\Bigl(\frac{f_{\pi}}{m_{\pi}}\Bigr)^{2}{\vec\tau}_1{\vec\tau}_2({\bf\Sigma}_1{\bf\nabla}_1)
({\bf\Sigma}_2{\bf\nabla}_2)D_{\pi}({\bf r}_1,{\bf r}_2),
\end{equation}
where $g_{\rho}$ and $f_{\pi}$ are the coupling strengths,
$D_{\rho}$ and $D_{\pi}$ are the meson propagators and ${\bf\Sigma}$
is the generalized Pauli matrix \cite{PNVR.04}. The Landau-Migdal
term $V_{\delta\pi}$ is the contact part of the nucleon-nucleon
interaction responsible for the short-range repulsion:
\begin{equation}
V_{\delta\pi}(1,2) =
g^{\prime}\Bigl(\frac{f_{\pi}}{m_{\pi}}\Bigr)^2{\vec\tau}_1{\vec\tau}_2{\bf\Sigma}_1{\bf\Sigma}_2
\delta({\bf r}_1 - {\bf r}_2),
\end{equation}
where the parameter $g^{\prime}$ = 0.6 is adjusted to reproduce
experimental data on the excitation energies of the Gamow-Teller
resonance in $^{208}$Pb and kept fixed in the calculations for other
nuclei, relying on the results obtained in Ref. \cite{PNVR.04}
within the relativistic QRPA. The amplitude $\Phi(\omega)$ describes
the coupling of the nucleons to vibrations generated by the coherent
nucleonic oscillations. In the time blocking approximation it has
the following operator form:
\begin{equation}
\Phi(\omega) = \sum\limits_{\mu,\eta}g_{\mu}^{(\eta)\dagger}{\tilde
R}^{0(\eta)}(\omega - \eta\omega_{\mu})g_{\mu}^{(\eta)}, \label{phi}
\end{equation}
where the index $\mu$ numerates vibrational modes (phonons) with
frequencies $\omega_{\mu}$ and generalized particle-vibration
coupling (PVC) amplitude matrices $g_{\mu}^{(\eta)}$, and the index
$\eta = \pm 1$ denotes forward and backward components, in full
analogy with the neutral-channel case \cite{LRT.07,LRT.08}.
%The
%energy-dependent effective interaction of Eq. (\ref{phi}) has the
%same form as the interaction of Eq. (\ref{phiphc0}), but the pairs
%of the channel indices ${k_1k_3}$ and ${k_2k_4}$ have different
%isospins.
This amplitude is responsible for the spreading mechanism caused by
the coupling between the ph and ph$\otimes$phonon configurations. In
the calculations presented below, the phonon space is truncated by
the angular momenta of the phonons at $J^{\pi} = 6^+$ and by their
frequencies at 15 MeV. The ph$\otimes$phonon configurations are
included up to 30 MeV of the excitation energy. The truncation is
justified by the subtraction of the term $\Phi(0)$ in Eq.
(\ref{inter}). This subtraction removes double counting of the PVC
effects from the residual interaction, guarantees the stability of
the solutions for the response function and provides faster
convergence of the renormalized PVC amplitude $\Phi(\omega)-\Phi(0)$
with respect to the phonon angular momenta and frequencies.
%This
%technique is discussed in detail in Ref. \cite{Tse.13}.
%, because the parameters of the density functional
%have been adjusted to experimental data for ground states and,
%therefore, include the particle-phonon correlations in the static
%approximation.

\begin{figure}
\centering
\includegraphics[width=12cm]{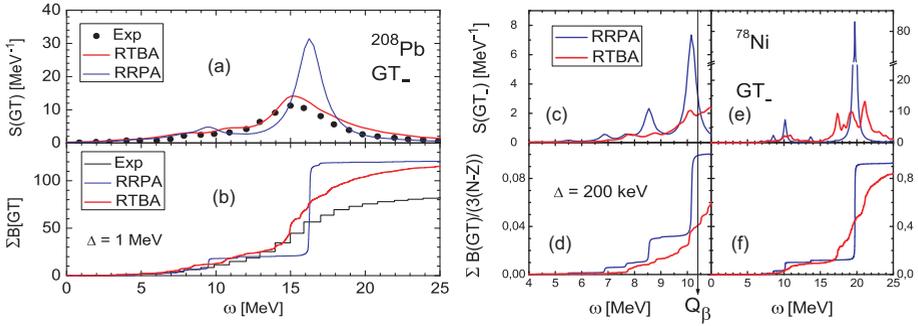}
\caption{Theoretical and experimental Gamow-Teller strength
distributions in $^{208}$Pb (a) and their cumulative sums (b). GTR
strength in the neutron-rich $^{78}$Ni: the low-energy portion of
the strength relevant for the beta-decay lifetime (c), its
cumulative sum (d), the overall strength distribution up to 25 MeV
(e) and its cumulative sum (f). $Q_{\beta}$ shows the beta-decay
window. }
\label{fig-1}       % Give a unique label
\end{figure}

The strength function $S^P(\omega)$ gives the spectral distribution
of the nuclear response for a particular external field $P$
expressed by the Gamow-Teller lowering operator:
\begin{equation}
S^{P}(\omega )=-\frac{1}{\pi} Im \langle P^{\dagger}R(E+i\Delta)P
\rangle, \ \ \ \ \ \ \ \ \ \ \ \ \omega = E + i\Delta, \ \ \ \ \ \ \
\ \ \ \  P_{GT_-} =
\sum\limits_{i=1}^{A}\tau_{-}^{(i)}{\bf\Sigma}_i.
\label{strf}%
\end{equation}
The finite imaginary part of the frequency argument is typically set
equal to the energy resolution of the corresponding experimental
data, in order to make a correct comparison.
%
%%%%%%%%%%%%%% 208-Pb
Figure \ref{fig-1}(a,b) shows the results for the GTR in $^{208}$Pb
obtained within the RRPA (without the last two terms in Eq.
(\ref{inter})) and RTBA (with the full interaction (\ref{inter})),
compared to the data of Ref. \cite{WOD.12}.
% , and that
%bab added
%the QRPA cannot account for the observed spreading width.
%For one-column wide figures use syntax of figure~\ref{fig-1}
The RRPA calculation produces a strength with the major peak at 16.5
MeV and a low-energy peak structure around 10 MeV. The exact Ikeda
sum rule is accommodated within the model space of pn-configurations
between -1800 MeV and 100 MeV, so that ~8\% of the $B(GT_-)$ is at
large negative energies because of the transitions to the Dirac sea
\cite{PNVR.04}. While the RRPA does not account for spreading
effects, within the RTBA the GTR acquires the spreading width
because of the coupling between the ph and ph$\otimes$phonon
configurations. Thus, the additional 5\% of the sum rule goes above
the considered energy region, while the total $B(GT_{+})$ is equal
to 0.34. Comparison to data shows that the spreading effects which
are taken into account in the RTBA are reproduced very well. The
overall strength is visibly quenched compared to the RRPA one,
however, the cumulative sum shows that the experimental integral
strength is still smaller than the RTBA one.
%The data from Ref. \cite{WOD.12} are shown without
%experimental error bars which are relatively small ($\pm$ 5\%), but
%even though the quenching obtained in the RTBA is only $\approx 0.5$
%of the observed one below 25 MeV.
A study of Ref. \cite{MMPV.12} has revealed that the rest of the
observed quenching can come from the finite momentum transfer, which
is not included in the present calculations. Figure \ref{fig-1}(c-f)
shows the GTR in neutron-rich $^{78}$Ni, whose properties are
interesting to study because of their astrophysical importance. The
damping caused by the particle-phonon coupling affects not only the
giant resonance, but also the low-lying strength, whose portion
below the beta-decay window $Q_{\beta}$ defines the lifetime. Within
the RTBA, this portion becomes considerably smaller than in the RRPA
and allows for direct determination of the half-life without
introducing an artificial quenching factor.
\section{Model space and quenching of the GTR}

In the context of the quenching problem the size of the basis
spanning the ph$\otimes$phonon configurations plays an important
role. In the numerical calculations, the energy window containing
these configurations is that, in which the amplitude of Eq.
(\ref{phi}) has non-zero matrix elements. The truncation of this
energy window is just a technical necessity and varies depending on
the available computing resources. In the case of  $^{208}$Pb it
turns out that almost the whole effect of coupling to the
ph$\otimes$phonon configurations appears within 30 MeV of the
excitation energy. An extension of the ph$\otimes$phonon
configuration window to 60 MeV does not influence the results. The
particle-hole proton-neutron configurations are included, however,
up to 100 MeV, so that about 10 units of the Ikeda sum rule are
absorbed by the GT strength between 25 and 100 MeV. Further increase
of the ph-basis size would spread the GT strength wider, but very
slowly, so that, although effectively there is no need to extend the
basis beyond 100 MeV, in principle, it is not clear how to justify
the cutoff rigorously.
%
%%%%%%%%%%%%%% 28-Si
%
Overall, the extension of the ph$\otimes$phonon window involves the
inversion of larger matrices, but nevertheless, in the present
formulation this window is considerably larger than that, for
instance, in typical shell-model calculations. Moreover, the present
approach allows an investigation of the convergence of the results
with respect to this window. As discussed above, in the case of
$^{208}$Pb the value of 30 MeV can be taken as the cutoff energy for
the ph$\otimes$phonon configurations.

\begin{figure}
\centering
\includegraphics[width=12cm]{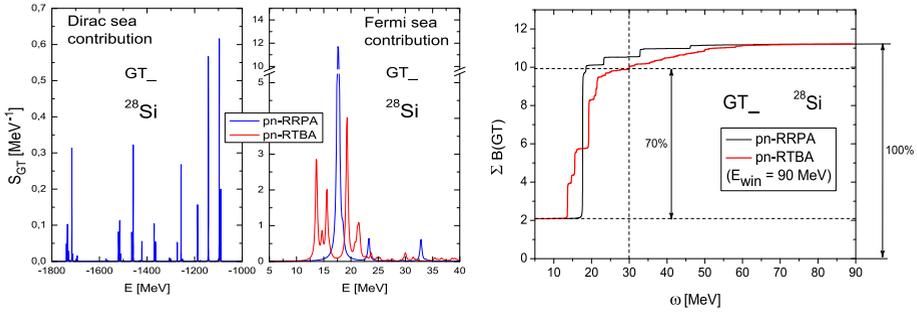}
\caption{Gamow-Teller strength distribution for $^{28}$Si. Left
panel: negative-energy part of the strength originated from the
proton-antineutron transitions to the Dirac sea. Middle:
positive-energy part of the GTR strength in pn-RRPA and in pn-RTBA.
Right: Running sums for the pn-RRPA and pn-RTBA strength
distributions.}
\label{fig-2}       % Give a unique label
\end{figure}

As the next example, a neutron deficient N=Z nucleus $^{28}$Si is
chosen. Being a considerably lighter than $^{208}$Pb, $^{28}$Si has
much lower level density, so that one can expect a higher saturation
energy for the ph$\otimes$phonon configurations. Indeed, the pn-RTBA
calculations confirm this expectation which is illustrated in Figure
\ref{fig-2}.
In this calculations the ph$\otimes$phonon configurations are
accomodated within 90 MeV energy interval, which is close to that,
where 1p1h pn-RRPA configurations are included (100 MeV). One can
see from the left panel of Figure \ref{fig-2}, that a part of the
GT$_-$ strength appears at large negative energies, which is the
typical feature of relativistic approaches using no-sea
approximation. In the nucleon sector, the cumulative GT$_-$ sum in
the pn-RTBA meets that obtained within the pn-RRPA at $E_c\sim$60
MeV, which is sufficiently smaller than the pn-RTBA window. Thus,
the GT$_-$ strength between 0 and 30 MeV (the energy region which is
usually associated with the GTR) amounts 70\% of the total GT$_-$
strength. This characterizes the ability of the pn-RTBA to explain
the quenching, which is observed experimentally in restricted energy
intervals, by the damping through 1p1h$\otimes$phonon
configurations. The GT$_+$ strength distribution (not shown) in this
nucleus looks very similar although not exactly the same as the
$GT_-$. The unique feature of N=Z nuclei is that the Ikeda sum rule
\begin{equation} S(GT_-) - S(GT_+) = 3(N-Z) \end{equation} is exactly zero which allows a
stringent test of the approach and of the numerical implementation.
Although the sum rule gives no values for the GT$_+$ and GT$_-$
distributions separately, it is fulfilled in the present
calculations for the GTR in $^{28}$Si with $\sim$0.2\% accuracy,
with respect to the total GT$_+$ or GT$_-$ strength.
%
%%%%%%%%%%%%%%%%%
%An extended discussion of the spin-isospin response calculated
%within pn-RTBA is presented in Ref. \cite{LBFMZ.14}. The pn-RTBA
%results are compared in detail to QRPA with G-matrix effective
%interaction and the large-basis shell model. All three models are
%applied to GTR in $^{132}$Sn nucleus allowing, for the first time, a
%comprehensive comparative study.
%
%%%%%%%%%%%%%%%%%
%
\section{Summary and outlook}

This paper presents selected results obtained in the covariant
nuclear field theory, recently developed for the nuclear response in
the isospin channel. The one-boson exchange nucleon-nucleon
interaction includes effective mesons and the emergent vibrational
quanta which account for the time dependence of both meson and
nucleonic fields. The performance of the approach is illustrated in
the calculations for Gamow-Teller strength in $^{208}$Pb, $^{78}$Ni
and $^{28}$Si. It is shown how the fragmentation of the strength
appears due to the coupling to vibrations, in addition to Landau
damping. The quenching of the GTR because of the spreading to high
energies and its connection with the completeness of the model space
are discussed. Further advancements of the pn-RTBA, such as
implementation of the superfluid-type pairing and higher-order
correlations, will be performed as the next steps of the model
development. Experimental data on the spin-isospin strength
distributions in broad energy intervals are expected from future
measurements of various exotic nuclei at the rare isotope beam
facilities. Such data will provide decisive arguments to constrain
many-body coupling schemes of the R(Q)TBA as well as the underlying
nuclear effective interactions.
\section{Acknowledgements}
Fruitful discussions with P. Ring, V. Tselyaev, T. Marketin, R.
Zegers, A. Brown, and D.-L. Fang are gratefully acknowledged. This
work was supported by the NSF grants PHY-1204486 nd PHY-1404343.
%For two-column wide figures use syntax of figure~\ref{fig-2}
%\begin{figure*}
%\centering
%\vspace*{5cm}       % Give the correct figure height in cm
%\label{fig-2}       % Give a unique label
%\end{figure*}

%For figure with sidecaption legend use syntax of figure
%\begin{figure}
%% Use the relevant command for your figure-insertion program
%\centering
%\sidecaption
%\includegraphics[width=5cm,clip]{tiger}
%\label{fig-3}       % Give a unique label
%\end{figure}

%For tables use syntax in table~\ref{tab-1}.
%\begin{table}
%\centering
%\label{tab-1}       % Give a unique label
% For LaTeX tables you can use
%\begin{tabular}{lll}
%\hline
%first & second & third  \\\hline
%number & number & number \\\hline
%\end{tabular}
% Or use
%\vspace*{5cm}  % with the correct table height
%\end{table}
%
% BibTeX or Biber users please use (the style is already called in the class, ensure that the "woc.bst" style is in your local directory)
% \bibliography{name or your bibliography database}

\end{document}